\documentstyle[aps,prl,psfig]{revtex}
\begin{document}
\draft
\title
{Nuclear Saturation with in-Medium Meson Exchange Interactions}

\author
{ R. Rapp$^{1}$, R. Machleidt$^{2}$, J.W. Durso$^{1,3}$, and G.E. Brown$^{1}$}
 
\address{
 $^1$Department of Physics, SUNY at Stony Brook, Stony Brook NY 11794-3800,
    U.S.A. \\
 $^2$Department of Physics, University of Idaho, Moscow, ID 83844, U.S.A. \\  
 $^3$Physics Department, Mount Holyoke College, South Hadley,
    MA 01075, U.S.A.\\ }
\date{May 1997}
\maketitle
\begin{abstract} 
\begin{center}
{\bf Abstract} 
\end{center} 
We show that the assumption of dropping meson masses
together with conventional many-body effects,
implemented in the relativistic Dirac-Brueckner formalism,
explains nuclear saturation.
We use a microscopic model for correlated $2\pi$ exchange and include
the standard many-body effects on the in-medium pion propagation,
which initially increase the attractive nucleon-nucleon ($NN$) potential
with density.
For the vector meson exchanges in both the $\pi\pi$ and $NN$ sector,
we assume Brown-Rho scaling
which---in concert with `chiral' $\pi\pi$ contact interactions---reduces
the attraction at higher densities. 
\end{abstract}

\twocolumn

The description of the ground state properties of nuclear matter in 
terms of an underlying microscopic nucleon-nucleon ($NN$) interaction has 
been one of the major challenges in modern nuclear physics 
(for a review see, {\it e.g.},~\cite{Mach89}). Nonrelativistic many-body 
approaches, such as Brueckner-Hartree-Fock (BHF),
resulted in the famous Coester band~\cite{Coe70}, which 
does not meet the empirical saturation point.
However, including the effective in-medium nucleon mass
in the relativistic Dirac structure of the nucleon,
as done in the so-called 
Dirac-BHF (DBHF) approach~\cite{Mach89,BM90},
generates additional saturation, 
resolving the aforementioned discrepancy. In these frameworks, the 
mesonic degrees of freedom are usually `frozen'; {\it i.e.}, no medium 
effects are applied to mesons. 
This is a natural, but unjustified, simplification.    

In the next simplest scenario, in analogy to the in-medium nucleon
mass (or by arguments based on scale invariance~\cite{BR91}),  one would 
assume that also the masses of the  
exchanged mesons (except for the pion, which is protected by its 
Goldstone nature) depend on the nuclear density. However, if the meson 
masses decrease at the same rate,  saturation cannot 
be reached at the empirical density, since the enhanced attraction 
in the effective $\sigma(550)$ 
exchange overwhelms the corresponding increase in repulsion due to 
$\omega(782)$ exchange~\cite{AmTj,BMP,BrMa}. 

More microscopically, the $\sigma(550)$ is understood as a 
correlated pair of pions in a relative s-wave (the strong attraction 
in the s-wave 
$\pi\pi$ interaction being chiefly due to $t$-channel $\rho$ exchange).
Therefore, as a consequence of the well established p-wave pion 
polarization in nuclear matter due to $\Delta$-hole and nucleon-hole 
excitations, the spectral distribution of the `effective' $\sigma$ 
meson undergoes an appreciable reshaping: the softening of the in-medium 
pion dispersion relation leads to a considerable shift of strength to  
lower energies~\cite{ARCSW,RDW1}. This, in turn, affects the $NN$ 
interaction in the nuclear environment, causing a marked increase in 
attraction. Qualitatively, this effect is comparable to that of a reduced 
`$\sigma(550)$' mass in the one boson exchange picture of the $NN$ 
potential.
However, additionally accounting for a density dependence in a 
chirally symmetric $\pi\pi$ interaction will
slow down the increase of attraction, particularly
at densities above $\rho_0$, thus improving saturation.

In this note we will show that a unified treatment of 
`conventional' many-body effects coupled with the assumption of 
decreasing meson masses, 
implemented in a DBHF framework, does indeed lead to reasonable 
saturation properties of nuclear matter.

First let us briefly review our model for the correlated two-pion 
exchange in the nucleon-nucleon potential~\cite{KDH,SHSPD}.
We employ a chirally improved version~\cite{RDW1} of an 
earlier meson exchange model~\cite{LDHS} for the free 
$\pi\pi$ interaction. Its dominant contribution at low 
energies stems from $t$-channel exchange of $\rho$ mesons which, in 
the scalar-isoscalar (`$\sigma$') channel, provides sufficient 
attraction to form the broad resonance-like structure around 
$E_{\pi\pi}$$\simeq$500~MeV. However the inclusion of 
$\pi\pi$ contact interactions, dictated by chiral symmetry~\cite{We66} 
and repulsive in nature, turned out to be crucial in avoiding unrealistic
$\pi\pi$ bound states in the 
nuclear environment~\cite{ARCSW,RDW1} while still enabling a satisfactory 
description of the free $\pi\pi$ scattering data~\cite{RDW1,RPhD}.  
The $\pi\pi$ interaction is then used to calculate a 
scattering amplitude for the $N\bar N \to \pi\pi \to N\bar N$ 
reaction~\cite{KDH,SHSPD}.  Schematically,  
\begin{eqnarray}
M_{N\bar N}(t') & = &
\tau_{B}^* \ G_{\pi\pi}^0 \tau_{B} \ +\tau_{B}^* \ G_{\pi\pi}^0 \ 
M_{\pi\pi} \ G_{\pi\pi}^0 \tau_{B}
\nonumber\\
& \equiv & M_{N\bar N}^{bare}(t') 
+ M_{N\bar N}^{rescat}(t') \ , 
\label{mnnb}
\end{eqnarray}
where $\tau_B$ denotes the transition Born amplitudes for 
$N\bar N\to\pi\pi$ (consisting of nucleon and $\Delta$ exchange)
and $M_{\pi\pi}$, $G_{\pi\pi}$ the $\pi\pi$ scattering amplitude 
and 2$\pi$ propagator, respectively. In the so-called pseudophysical 
region ({\it i.e.}, for energies $\sqrt{t'}$ well below the 
nucleon-antinucleon threshold), it is sufficient to consider 
$\pi\pi$/$K\bar K$ intermediate states (contained in our model for
$M_{\pi\pi}$). Again, the full transition amplitudes, 
\begin{equation}
\tau=\tau_B + \tau_B \ G_{\pi\pi} \ M_{\pi\pi} \ ,  
\end{equation}
are in good agreement with quasiempirical information obtained by  
H\"ohler {\it et al.}~\cite{Hoeh}. Finally, the $I$=$J$=0 $N\bar N$ 
amplitude is related via a dispersion relation to the 
correlated $2\pi$ exchange potential in the $NN$ channel as 
\begin{equation}  
V_{NN}^{2\pi,corr}(t)=-\frac{1}{\pi} \int\limits_{4m_\pi^2}^{t_c}
dt' \ \frac{\eta_{00}(t')}{t'-t},  
\label{VNNt} 
\end{equation} 
\begin{equation}
\eta_{00}(t') = \frac{4m_N^2}{t'-4m_N^2} \ \frac{3}{4\pi} \ 
Im M_{N\bar N}^{00,rescat}(t'),
\label{eta00m}
\end{equation}
where $t=(k'-k)^2\equiv k''^2$ is the 4-momentum 
transfer between the 
two  nucleons. For practical purposes,  
we extract a coordinate space potential by taking the 
quasistatic limit, $k''\simeq -\vec{k''}$, and  Fourier transforming 
(\ref{VNNt}):
\begin{eqnarray}  
V_{NN}^{2\pi,corr}(r) = -\frac{1}{\pi}\int\limits_{4m_\pi^2}^{t_c} dt' 
 \  \eta_{00}(t') \ \frac{\exp (-\sqrt{t'}r)}{r} \ .  
\label{vnnsr}
\end{eqnarray} 

Let us now turn to the medium modifications. In order to preserve  
the analogy with the one boson exchange
picture of the $NN$ interaction, we regard the correlated 2$\pi$ 
contribution as an `effective $\sigma$'.  
That means that we restrict the effects of the nuclear environment 
in Eq.~(\ref{mnnb}) to the $\pi\pi$ amplitude $M_{\pi\pi}$. The latter
is obtained by solving a Lippmann-Schwinger-type equation
\begin{equation}
M_{\pi\pi} = V_{\pi\pi} + V_{\pi\pi} \ G_{\pi\pi} \ M_{\pi\pi}  \ , 
\end{equation} 
where medium effects are induced in both $G_{\pi\pi}$ (through a change 
in the pion dispersion relation) and in the $\pi\pi$ interaction kernel 
$V_{\pi\pi}$. As has been shown in Ref.~\cite{RDW2}, the restriction to 
medium effects in $G_{\pi\pi}$ yields $NN$ potentials which will not be 
compatible with nuclear saturation. However, as was conjectured in  
Ref.~\cite{RDW2}, the inclusion of density-dependent meson masses 
and coupling constants in the $\pi\pi$ interaction kernel might provide
a remedy to this problem. Here we work at mean field level, changing
only $f_\pi^*$ and masses according to 
\begin{equation} 
\Phi(\rho)=(1-C \ \rho/\rho_0)=\frac{f_\pi^*}{f_\pi}=\frac{m^*}{m}
=\frac{\Lambda^*}{\Lambda} \qquad  {\rm etc.}, 
\end{equation} 
with the scaling factor taken to be $C=0.15$ (which is in line with 
QCD sum rule analyses~\cite{HaLe}, and has become known
as Brown-Rho (BR) scaling~\cite{BR91}).  
Explicitly, the following quantities in the $\pi\pi$ interaction 
kernel are subject to this medium dependence (indicated by an asterisk): 
\begin{itemize} 
\item[(i)] the mass $m_\rho^*$ and formfactor cutoff $\Lambda_\rho^*$ of 
the $t$-channel $\rho$ exchange; 
\item[(ii)] the pion decay constant $f_\pi^*$ and $m_\rho^*$, both entering 
the $\pi\pi$ contact interactions.  
\end{itemize} 
Since the KSFR relation is supposed to remain valid in the medium, 
\begin{equation}
2g_{\pi\pi\rho}^2 (f_\pi^*)^2=(m_\rho^*)^2 \ ,
\label{ksfr}
\end{equation}
the $\rho\pi\pi$ coupling constant is not affected. The same holds for 
masses and cutoffs related to the pion due to its Goldstone boson nature.
For consistency, corresponding modifications are also applied in 
the single-pion self energy, $\Sigma_\pi$, which is responsible for the 
many-body effects in the (uncorrelated)
2$\pi$ propagator $G_{\pi\pi}$. It is calculated in terms of standard p-wave
particle-hole ($NN^{-1}$ and $\Delta N^{-1}$) excitations~\cite{MSTV}: 
\begin{equation} 
\Sigma_\pi(\omega,k) = -z_\pi^2 \ k^2 \ \chi(\omega,k) \ . 
\label{sigpi} 
\end{equation}    
The pion susceptibility, $\chi$, contains short-range correlation effects
between particle and hole parametrized by Migdal parameters $g'$. Since the 
dynamical origin of the latter is partly due to vector meson exchange we 
employ a density dependent form 
\begin{eqnarray}
(g'_{NN})^* & = & 0.65+C \ \rho / \rho_0
\nonumber\\
(g'_{N\Delta})^* & = & (g'_{\Delta\Delta})^*
 =  0.35+C \ \rho / \rho_0 \ ,
\label{gprime}
\end{eqnarray}
which is somewhat weaker than that suggested in Ref.~\cite{BBLW}, where it
was erroneously assumed that the anomalous $\rho NN$ coupling through 
$\kappa_V$ should be enhanced by the factor $(m_N/m_{N^*})^2$; in 
fact one can show from the equations of motion that only the 
convection current contribution should be. 
We also use effective masses for nucleons and deltas: 
\begin{eqnarray}
m_N^* & = & m_N(1-C \ \rho/\rho_0)
\nonumber\\
m_\Delta^* & = & m_\Delta-(m_N-m_N^*) \ .  
\end{eqnarray}
The factor $z_\pi$ in Eq.~(\ref{sigpi}) accounts for effects of 
$\pi N$ s-wave interactions~\cite{MSTV} which, in a scattering length 
approximation, is estimated to be  
\begin{equation} 
z_\pi=(1+4\pi\lambda\rho)^{-1/2} 
\end{equation} 
with $\lambda$$\simeq$0.11~fm$^3$.

\begin{figure}[t]
\vspace*{-7.5mm}
\psfig{figure=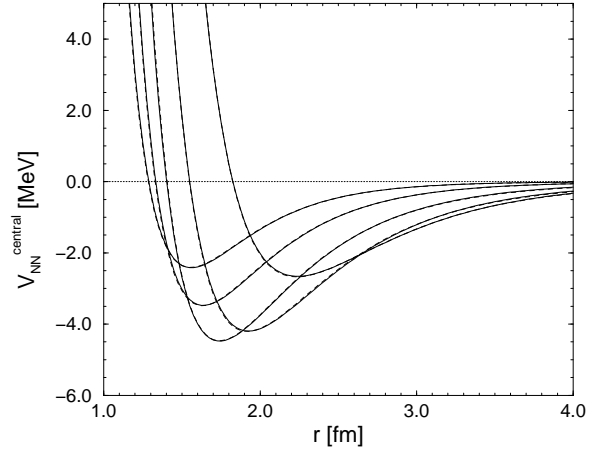,height=7cm,angle=-90}
\vspace*{0.5cm}
\caption{In-medium central $NN$ potential, Eq.\ (5), supplemented
by $\omega$ exchange (using $m^*_\omega/m_\omega=\Phi(\rho)$)
for various nuclear matter densities. The solid lines are our 
microscopic results from in-medium correlated 2$\pi$ exchange and
the dashed lines correspond to a fit with two zero-width sigma 
mesons with the parameters given in Table~I. The densities are 
0, $0.5\rho_0$, $\rho_0$, $1.5\rho_0$ , and $2\rho_0$ from top
to bottom at the right side of the graph.}
\end{figure}

The resulting in-medium central $NN$ potential, Eq.~(\ref{vnnsr}),
supplemented with zero-width $\omega(782)$ exchange (using 
$m_\omega^*/m_\omega=\Phi(\rho)$), is shown in Fig.~1. At low densities
the increase in attraction in
$V_{NN}^{2\pi,corr}(r)$ dominates since both the softening of the pion 
dispersion relation and the enhanced attraction due to $t$-channel $\rho$ 
exchange with reduced mass result in a marked shift of spectral strength 
of the `$\sigma$' ({\it i.e.} in $\eta_{00}(t')$) to lower energies. 
However, at higher densities this tendency becomes more and more suppressed:
on the one hand, both the increasing $g'$ and $z_\pi$ decelerate the pion 
softening, and on the other hand the contact interactions in the $\pi\pi$
kernel $V_{\pi\pi}$, being proportional to $(f_\pi^*)^{-2}$, balance the 
increase in attraction in $\rho$ exchange. As a result, the attraction from  
$V_{NN}^{2\pi,corr}(r)$ increases less rapidly than the repulsion in
$V_{NN}^\omega(r)$ (due to a reduced $\omega(782)$ mass).  That means that
at low densities our microscopically calculated $2\pi$ exchange potential
behaves approximately like a scalar meson, with mass decreasing at the same
rate as for the $\omega$.  Approaching $\rho_0$, the contact terms dictated
by chiral symmetry slow the decrease of the effective $\sigma$ mass, thus
enabling saturation.


To investigate the quantitative impact on nuclear matter properties, 
we apply the in-medium $2\pi$-exchange 
\begin{table}[t]
\caption{Density-dependent 
parametrization of the correlated in-medium $2\pi$ exchange in terms
of two zero-width sigma bosons ($\Lambda_\sigma = 3$ GeV in all cases).}
\begin{tabular}{ccccc}
density $(\rho_0)$ 
     &  $m_{\sigma_1}$ (MeV)
     &  $g^{2}_{\sigma_1}/4\pi$
     &  $m_{\sigma_2}$ (MeV)
     &  $g^{2}_{\sigma_2}/4\pi$
\\ \hline
    0        &       640    &    4.36      &       377    &    0.640   \\      
    0.5      &       538    &    3.405     &       295    &    0.450   \\
    0.75     &       471    &    2.875     &       249    &    0.309   \\ 
    1.0      &       427    &    2.53      &       220    &    0.264   \\ 
    1.25     &       395    &    2.35      &       200    &    0.234   \\
    1.5      &       365    &    2.17      &       190    &    0.212   \\
    2.0      &       315    &    1.765     &       181    &    0.162   \\
\end{tabular}
\end{table}
\noindent 
model in a relativistic DBHF calculation. For this purpose,
we start from the Bonn-B potential~\cite{foot} which is
a one-boson-exchange potential (OBEP) that includes the
$\pi$,  $\eta$, $\rho(770)$, $\omega(782)$, and
$a_0(980)$ mesons as well as a zero-width $\sigma(550)$ boson
and describes free-space $NN$ scattering accurately up to
pion-production threshold.
In this OBEP, we replace the zero-width $\sigma$
by our microscopic model of $2\pi$ exchange explained
above. To make the calculations more
tractable, we parametrize
$V_{NN}^{2\pi,corr}(r;\rho)$ in terms of
two sharp scalar mesons with density-dependent masses and 
coupling constants, {\it cf.} Table~I and dashed lines in Fig.~1. 
As discussed in the literature~\cite{DJV80,MHE87}, 
the correlated $2\pi$ exchange accounts for more than half
of the intermediate-range attractiont.  The remainder is provided by
uncorrelated $2\pi$ exchange which we parametrize, in our present
calculations, in terms of a `rest' $\sigma$ boson with less than half
the coupling strength of the full $\sigma$  of a typical OBEP.
This combination, plus the five mesons mentioned above
({\it cf.}\ Table~II) reproduces free-space $NN$ scattering as well as
the original Bonn-B potential.

This model for the $NN$ interaction is now applied to nuclear
matter. We use the DBHF approach~\cite{Mach89,BM90}, scale
the nucleon mass as well as the meson and cutoff masses 
of $\rho(770)$ and $\omega(782)$ according to Eq.\ (7), and use the 
density-dependent correlated $2\pi$ exchange.  Our result for
the energy per nucleon as a function of density is shown
in Fig.~2 by the solid curve.  The predicted saturation  
energy is --15.1 MeV/nucleon at a density corresponding 
to $k_F = 1.32$ fm$^{-1}$ 
($\rho=2k_F^3/3\pi^2$),
in good agreement with the empirical values.
The incompressibility, $K_V^{-1}$, is 356 MeV, which is
slightly above the currently favored 
\begin{table}[b] 
\caption{Meson parameters for the one-boson-exchanges involved
in the $NN$ model applied in the present work.}
\begin{tabular}{cccccc}
     & $J^P$ & $I$ 
     & $m_{\alpha}$ (MeV) 
     &  $g^{2}_{\alpha}/4\pi$
     &  $\Lambda_{\alpha}$ (GeV) \\
\hline

$\pi$ & $0^-$ & 1 & 138.03  & 14.6 & 1.2   \\

$\eta$ & $0^-$ & 0 & 548.8  &   5  & 1.5   \\

$\omega$ & $1^-$ & 0 & 782.6 & 20   &  1.5   \\

$\rho$ & $1^-$ & 1 & 769    &  0.95$^*$ & 1.3   \\

$a_0$ & $0^+$ & 1 & 983  & 2.9908  &   1.5  \\

rest-$\sigma$ & $0^+$ & 0 & 517 & 2.8386  &  3.0  \\
\end{tabular}
$^*$ $f_\rho/g_\rho=6.1$
\end{table}

\begin{figure}[h]
\vspace*{0.5cm}
\psfig{figure=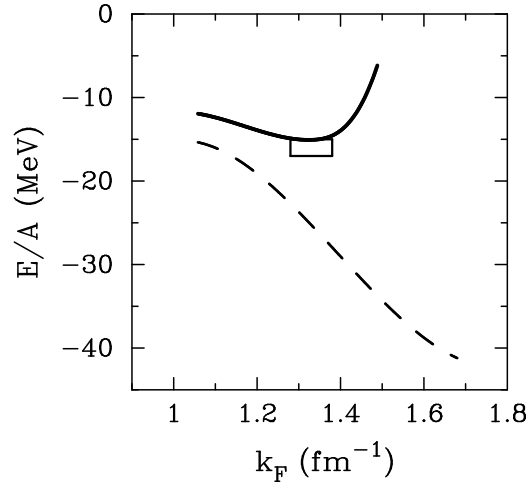,height=8cm}
\vspace*{-2cm}
\caption{Energy per nucleon in nuclear matter, $E/A$, as a function
of density in terms of the Fermi momentum $k_F$.
The solid curve displays our main result which is obtained
from a DBHF calculation
applying BR scaling, Eq.\ (7), to $\rho$ and $\omega$
and using the correlated in-medium $2\pi$ exchange 
explained in the text.
When the latter is replaced by a zero-width $\sigma$ boson
which is subjected to BR scaling, then the dashed curve is
obtained.  The box represents the empirical values for nuclear matter 
saturation ($E/A=-16\pm 1$ MeV, $k_F=1.33\pm 0.05$ fm$^{-1}$).}
\end{figure}

\noindent
range.  (Similar results have been obtained with a more schematic
mean field model~\cite{SBMR} in which introduction of more rapid
scaling of the nucleon effective mass,
$m_N^*/m_n=\sqrt{g_A^*/g_A}\Phi(\rho)$, with loop effects~\cite{BR91},
substantially decreased $K_V^{-1}$.  We believe that it would do so in
our case too, but because of greatly increased technical complications,
we have not yet been able to go to higher loop order.)

Our results may be contrasted with a simpler calculation in which
a conventional OBEP is applied
and the $\sigma$ boson (besides $\rho$ and $\omega$) is
subjected to BR scaling. As discussed, no saturation
can be achieved in this case, as clearly revealed
by the dashed curve in Fig.~2.

In summary, we have studied the impact of medium modifications
in meson exchanges of the $NN$ potential on saturation properties 
of nuclear matter. The two-pion exchange contribution has been calculated
microscopically with the standard many-body effects on the in-medium 
pion propagation taken into account. For the vector meson exchanges in both
the $\pi\pi$ and the nucleon-nucleon sector we assumed a universal 
decrease in their masses $m_{\rho,\omega}$, which has also been applied 
to the nucleon mass in the many-body parts of the calculation. 
We find that in a relativistic DBHF calculation this provides
a reasonable saturation mechanism, thus improving earlier results in which 
a universal scaling of the exchanged meson masses 
(including the mass of an alleged `$\sigma$ meson')
could not achieve a stable configuration. The main reason for this difference 
is the fact that the combination of nuclear many-body effects on the pions 
and the short-range repulsive part of a (broken) chirally symmetric $\pi\pi$ 
interaction suppresses the increase of attraction in the $\sigma$ channel 
at nuclear densities above the empirical saturation value.  Thus the chiral
constraints on the $\pi\pi$ interaction in our model~\cite{RDW2} are seen 
to play a
subtle, but crucial, role in nuclear matter saturation.       
Whether the mass scaling of the vector mesons, which plays an equally
important role in the saturation process, can be understood in
a similar (microscopic) way to that of the `$\sigma$' remains a question 
for further investigation.

One of us (RR) acknowledges support from the Alexander-von-Humboldt 
foundation as a Feodor-Lynen fellow. 
This work was supported in part by the U.S. Department 
of Energy under Contract No.\ DE-FG02-88ER40388
and by the U.S. National Science Foundation 
under Grant No.\ PHY-9211607.

\end{document}